\documentclass[runningheads]{llncs}
\usepackage[T1]{fontenc}
\usepackage{amsmath}
\usepackage{newtxtext}
\usepackage[varvw]{newtxmath}
\usepackage{graphicx}
\usepackage{booktabs}
\usepackage{adjustbox}
\usepackage{cite}
\usepackage{url}

\usepackage{hyperref}
\usepackage{color}

\hypersetup{
  pdftitle={Empirical Evaluation of Machine Learning for Cloud Cryptojacking Detection: Effects of Preprocessing and Evaluation Protocols},
  pdfauthor={Amitabh Chakravorty},
  pdflang={en-US},
  hidelinks
}
\begin{document}
\title{Empirical Evaluation of Machine Learning for Cloud Cryptojacking Detection: Effects of Preprocessing and Evaluation Protocols}
\titlerunning{Machine Learning for Cloud Cryptojacking Detection}
\author{Amitabh Chakravorty \and Nelly Elsayed}
\authorrunning{A. Chakravorty and N. Elsayed}
\institute{School of Information Technology, University of Cincinnati, Cincinnati, OH, USA\\
\email{chakraa4@mail.uc.edu, elsayeny@ucmail.uc.edu}}
\maketitle
\begin{abstract}
Artificial intelligence (AI) has been widely applied to cloud cryptojacking detection. Classical machine learning models remain among the most widely used approaches for cloud cryptojacking detection. This paper presents an empirical evaluation of six widely used classical machine learning models for AI-based cloud cryptojacking detection under different preprocessing and evaluation settings.
Because no public cloud-native cryptojacking dataset is currently available, experiments were conducted using two publicly available proxy datasets, DS2OS and NSL-KDD.
The evaluated models include Random Forest, XGBoost, LightGBM, Gradient Boosting, Decision Tree, and K-Nearest Neighbors (KNN).
Experimental results show that preprocessing choices and evaluation protocols have a measurable impact on model performance. On DS2OS, removing identifier-related features reduced the accuracy of most models to approximately 96\%, while KNN achieved higher accuracy with lower attack recall. On NSL-KDD, models obtained 77 to 80\% accuracy on the official test set containing unseen attack types, compared with above 99.5\% under a random train-test split. Attack-class precision, recall, and Matthews Correlation Coefficient (MCC) further provide complementary insights into classifier performance beyond overall accuracy.
The results provide an empirical comparison of commonly used classical machine learning models under consistent experimental settings and offer insights into how preprocessing, dataset selection, and evaluation methodology influence reported performance.

\end{abstract}
%
%%%%%%%%%%%%%%%%%%%%%%%%%%%%%%%%%%%%%%%%%%%%%%%%%%%%%%%%%%%%%%%%%%%%%%%%%%%%%%%
\section{Introduction}
Cryptojacking is the unauthorized use of computing resources to mine cryptocurrency~\cite{tekiner2021sok}. In cloud environments the threat is acute. Elasticity, multi-tenancy and automated provisioning let an attacker scale a single foothold into large mining workloads while the cost lands on the victim~\cite{microsoft2023cryptojacking}. Every hijacked processor cycle also burns energy on a workload that serves no legitimate purpose, so timely detection supports sustainable cloud operation as much as it supports security. The behavior is stealthy by design~\cite{varlioglu2022dangerous,varlioglu2024pulse}. Mining processes throttle their own resource use and mimic legitimate high-intensity workloads such as model training or batch analytics, which makes them difficult to separate from benign activity~\cite{tekiner2021sok}.

A large body of work applies artificial intelligence to cryptojacking detection, including in cloud settings, using both classical machine learning and deep learning techniques~\cite{naseem2021minos,varlioglu2026detecting,sanda2023deep,ref10,ref7}. Reported performance is generally high across a variety of datasets and experimental settings. However, differences in datasets, preprocessing strategies, and evaluation protocols make direct comparison between studies difficult. An empirical evaluation under consistent experimental conditions is therefore valuable for understanding the relative performance of commonly used machine learning models.

To the best of our knowledge, no publicly available cloud-native cryptojacking dataset currently exists~\cite{chakravorty2026detecting}. This study therefore evaluates six widely used classical machine learning models using two publicly available proxy datasets, DS2OS and NSL-KDD, under consistent preprocessing and evaluation settings.

The main contributions of our proposed paper are:
\begin{itemize}
\item We empirically evaluate six widely used classical machine learning models for AI-based cloud cryptojacking detection on two publicly available proxy datasets.
\item We analyze the impact of preprocessing choices, including feature leakage removal, on model performance using the DS2OS dataset.
\item We compare model performance under different evaluation protocols using the official NSL-KDD holdout and a random train-test split.
\item We compare model accuracy, attack-class metrics, and computational cost to provide a comprehensive empirical assessment of the evaluated models.
\end{itemize}

Section~\ref{sec:bg} summarizes the background and motivation. Section~\ref{sec:method} presents the methodology. Section~\ref{sec:setup} describes the experimental setup. Section~\ref{sec:results} reports the empirical results. Section~\ref{sec:discussion} discusses the findings. Section~\ref{sec:threats} presents the threats to validity, and Section~\ref{sec:conclusion} concludes the paper.

%%%%%%%%%%%%%%%%%%%%%%%%%%%%%%%%%%%%%%%%%%%%%%%%%%%%%%%%%%%%%%%%%%%%%%%%%%%%%%%
\section{Background}
\label{sec:bg}

This section briefly summarizes the background relevant to the empirical evaluation conducted in this study, including the machine learning models commonly used for cloud cryptojacking detection and the publicly available datasets used for evaluation.

\subsection{Classical Machine Learning Dominates Practice}
Classical machine learning remains widely used for AI-based cloud cryptojacking detection because of its low computational cost, interpretability, and ease of deployment. Random Forest, Decision Tree, K-Nearest Neighbors, and boosting-based models are among the most frequently adopted approaches, including in prior studies on the DS2OS dataset used in this work~\cite{ref39,ref21}. Accordingly, this study focuses on these representative classical machine learning models.

\subsection{Datasets Are Opaque and Rarely Cloud-Native}
At present, no publicly available dataset specifically captures cloud-native cryptojacking behavior. Consequently, most empirical studies rely on publicly available intrusion detection datasets as behavioral proxies. Following this practice, this work uses the DS2OS and NSL-KDD datasets to provide a consistent empirical evaluation of the selected machine learning models.

%%%%%%%%%%%%%%%%%%%%%%%%%%%%%%%%%%%%%%%%%%%%%%%%%%%%%%%%%%%%%%%%%%%%%%%%%%%%%%%
\section{Research Methodology}
\label{sec:method}
The objective of this study is to empirically evaluate the performance of commonly used classical machine learning models for cloud cryptojacking detection under consistent preprocessing and evaluation settings. The evaluation considers different datasets, preprocessing strategies, and evaluation protocols to examine their impact on model performance.

The two datasets were selected because they support complementary evaluation scenarios of the empirical evaluation. DS2OS enables analysis of the impact of preprocessing choices, particularly the removal of identifier-related features that may influence model performance. NSL-KDD complements this analysis through its official train-test partition, which allows evaluation on previously unseen attack types. Together, the datasets provide a broader empirical assessment of model behavior under different preprocessing and evaluation conditions.

\subsection{Public Proxy Dataset Rationale and Scope Boundary}
Because no publicly available cloud-native cryptojacking dataset currently exists, this study uses two publicly available proxy datasets. DS2OS was selected to evaluate the impact of preprocessing choices, including feature leakage removal, while NSL-KDD was selected because its official train-test partition enables evaluation on unseen attack types. Although neither dataset represents real cloud telemetry, both have been widely used as behavioral proxies in intrusion detection research and provide a consistent basis for empirical comparison.

DS2OS and NSL-KDD were selected over other public intrusion detection datasets such as BoT-IoT and CICIDS because together they directly support this study's two evaluation scenarios. DS2OS exposes identifier-related feature leakage and NSL-KDD's official holdout enables evaluation on unseen attack types.

NSL-KDD represents network-level intrusion with 41 features describing connection metadata, protocol behavior, byte volumes and error statistics~\cite{nslkdd_dataset}. Cryptojacking generates detectable network signatures, including persistent outbound connections to mining pools, periodic Stratum exchanges and abnormal session durations~\cite{ref29,ref7}. More important for this study, the official \mbox{KDDTest+} holdout contains attack types absent from \mbox{KDDTrain+}. That property tests generalization to unseen threats, which is the exact condition a deployed detector faces~\cite{ref25}. Although NSL-KDD is a legacy intrusion detection dataset, its official train-test partition provides a standard benchmark for evaluating model generalization to previously unseen attack types~\cite{tavallaee2009nslkdd}.

What transfers across both datasets is distributed attack orchestration, realistic class imbalance and anomalous patterns embedded in dominant benign traffic~\cite{ref20,ref27}. What does not transfer is cloud-specific telemetry. Neither dataset includes Kubernetes pod metrics, Docker container resource limits, VM CPU steal time or cloud billing anomalies~\cite{ref10,ref32}. Although these datasets do not capture cloud-native telemetry directly, they provide representative behavioral characteristics that support a consistent empirical comparison of the evaluated machine learning models.

\subsection{Model Selection}
We evaluate six classical machine learning families that are widely reported in cryptojacking and intrusion detection studies: RF, XGBoost, LightGBM, Gradient Boosting, DT and KNN. Together they span the relevant accuracy and efficiency trade-offs.

Deep learning models were not considered because the objective of this study is to provide a consistent empirical evaluation of representative classical machine learning approaches that are widely adopted in the literature. Extending the evaluation to deep learning models remains future work.

\subsection{Evaluation Protocol and Metrics}
All models were evaluated using consistent train-test protocols to enable a fair empirical comparison. For DS2OS, a stratified 70/30 train-test split was adopted, while NSL-KDD was evaluated using both its official train-test partition and a random train-test split to compare performance under different evaluation settings.

To assess the stability of the reported results, each experiment was repeated using ten random seeds. The resulting standard deviation remained below 0.2 percentage points for DS2OS and below 1 percentage point for NSL-KDD, indicating that the reported performance is stable across runs.

Model performance was evaluated using accuracy, precision, recall, F1-score, Matthews Correlation Coefficient (MCC), and training time. MCC was included because it stays informative when classes are imbalanced~\cite{chicco2020mcc}. These metrics provide complementary measures of predictive performance, class-specific detection capability, and computational efficiency.

\subsection{Leakage Score}\label{subsec:defs}
Feature leakage, in which a feature reveals the label through an artifact of data collection rather than through behavior, is a recognized source of inflated results in machine learning research~\cite{kaufman2012leakage,kapoor2023leakage} and a documented pitfall in security applications~\cite{arp2022dos}. To identify potential feature leakage in the DS2OS dataset, each feature was evaluated independently using a simple lookup classifier. The resulting leakage score, denoted by $L(f)$, measures the predictive capability of a single feature with respect to the class label:
\begin{equation}
L(f)=\sum_{v\in V_f}\frac{n_v}{N}\max_{c\in C}\hat{p}(y=c|f=v)
\label{eq:leak}
\end{equation}

Because highly imbalanced datasets naturally produce large values of $L(f)$, we normalize the score with respect to the majority-class proportion:
\begin{equation}
L^{*}(f)=\frac{L(f)-p_{\mathrm{maj}}}{1-p_{\mathrm{maj}}}
\label{eq:leakadj}
\end{equation}
where $p_{\mathrm{maj}}$ denotes the majority-class proportion. The adjusted leakage score $L^{*}(f)$ ranges from 0 to 1, with values close to 1 indicating that a feature acts as a near-deterministic predictor of the class label. This score was used only as a preprocessing diagnostic to identify identifier-related features prior to model training and was not used for feature selection during model optimization.

%%%%%%%%%%%%%%%%%%%%%%%%%%%%%%%%%%%%%%%%%%%%%%%%%%%%%%%%%%%%%%%%%%%%%%%%%%%%%%%
\section{Experimental Design}
\label{sec:setup}

\subsection{Datasets}
\textbf{DS2OS.} The dataset contains 357,952 samples of IoT service traffic with 12 raw features and a target that distinguishes normal behavior from seven attack types~\cite{ds2os_dataset}.
Identifier-related features were removed during preprocessing based on the leakage analysis described in Section~\ref{subsec:defs}. The remaining five behavioral features were used for model training.

\textbf{NSL-KDD.} The dataset contains 125,973 training samples and 22,544 test samples with 41 features~\cite{nslkdd_dataset,tavallaee2009nslkdd}. We aggregated the original attack categories into a binary problem of normal versus attack. The official \mbox{KDDTest+} holdout, which includes attack types not present in \mbox{KDDTrain+} such as mscan, saint, apache2 and processtable, was used without modification to preserve its role as a generalization benchmark.

\subsection{Preprocessing Pipeline}\label{subsec:preprocessing}
All datasets passed through a standardized preprocessing pipeline in which categorical features were label encoded. For DS2OS we used a stratified 70/30 split with fixed random seed 42. For NSL-KDD we used the official \mbox{KDDTrain+} and \mbox{KDDTest+} partitions for the novel-attack holdout evaluation. A separate stratified 70/30 random split of \mbox{KDDTrain+} was used only for the seen-attack comparison reported in Table~\ref{tab:gengap}. The DS2OS training partition exhibited severe imbalance. The Synthetic Minority Over-sampling Technique (SMOTE)~\cite{chawla2002smote} was applied to the DS2OS training partition only. Preliminary tests without SMOTE produced models that predicted only the majority class, demonstrating that overall accuracy alone can mask poor attack detection performance on highly imbalanced datasets.

SMOTE was not applied to NSL-KDD, whose \mbox{KDDTrain+} distribution is approximately 53 percent normal and 47 percent attack, because oversampling a near-balanced set risks injecting synthetic noise without benefit. The \mbox{KDDTest+} holdout was never resampled, since resampling the test partition would defeat the generalization evaluation. All features were standardized to zero mean and unit variance,
\begin{equation}
z = \frac{x-\mu}{\sigma},
\label{eq:std}
\end{equation}
which is essential for the distance-based and gradient-based learners. The original imbalanced distribution was preserved in the DS2OS test set (104,381 normal versus 3,005 attack) to reflect realistic deployment where attacks are rare.

Because DS2OS contains categorical features, an additional sensitivity analysis using SMOTE-N produced negligible differences in model performance, indicating that the reported results are robust to the choice of oversampling method.

\subsection{Hyperparameters and Environment}
Hyperparameters were selected based on commonly reported configurations in the literature. Automated hyperparameter optimization was not performed in order to maintain a consistent comparison across the evaluated models. Table~\ref{tab:hyperparameters} summarizes the configuration used for each evaluated model. All experiments ran on a Google Colab runtime with two virtual CPUs, using Python 3.13, scikit-learn 1.6.1, XGBoost 3.4.1, LightGBM 4.6.0 and imbalanced-learn 0.14.2. Training times are wall-clock measurements from that runtime, so they support comparison between models within this study rather than absolute benchmarking.

\begin{table}[t]
\caption{Model configurations used in the experiments.}
\label{tab:hyperparameters}
\centering
\setlength{\tabcolsep}{5pt}
\begin{tabular}{@{}ll@{}}
\toprule
\textbf{Model} & \textbf{Configuration} \\
\midrule
Random Forest & 100 trees, maximum depth = 20, minimum samples split = 5 \\
Decision Tree & Maximum depth = 15, minimum samples split = 5 \\
Gradient Boosting & 100 trees, learning rate = 0.1, maximum depth = 5 \\
XGBoost & 100 trees, learning rate = 0.1, maximum depth = 10 \\
LightGBM & 100 trees, maximum depth = 10 \\
K-Nearest Neighbors & $k=5$ neighbors \\
\bottomrule
\end{tabular}
\end{table}

\subsection{Reproducibility Artifacts}
The preprocessing, training and analysis scripts that produce every table and figure in this paper are publicly available at \url{https://github.com/AmitabhCh822/cloud-cryptojacking-ml-evaluation}~\cite{chakravorty2026code}, together with the outputs of the run reported in this paper.

%%%%%%%%%%%%%%%%%%%%%%%%%%%%%%%%%%%%%%%%%%%%%%%%%%%%%%%%%%%%%%%%%%%%%%%%%%%%%%%
\section{Experimental Results}
\label{sec:results}

\subsection{Overall Performance}
Table~\ref{tab:validation_results} summarizes performance of the six evaluated machine learning models on the DS2OS and NSL-KDD datasets. The results show that model performance varies considerably across datasets because of differences in preprocessing and evaluation protocols.
On DS2OS, tree-based models converged on nearly identical accuracy of about 96.26 percent, with KNN reaching 99.21 percent. The convergence reflects the narrow feature space after leakage removal. With only five low-cardinality behavioral features, the tree-based algorithms discover essentially the same decision boundaries. On NSL-KDD, accuracy ranged from 76.76 percent for KNN to 80.47 percent for XGBoost.

The lower performance on NSL-KDD reflects the use of the official \mbox{KDDTrain+}/\allowbreak \mbox{KDDTest+} partition, which evaluates models on previously unseen attack types and therefore represents a more challenging evaluation scenario.

\begin{table}[t]
\caption{Performance of the evaluated machine learning models.}
\label{tab:validation_results}
\centering
\setlength{\tabcolsep}{5pt}
\begin{tabular}{@{}llcccc@{}}
\toprule
\textbf{Dataset} & \textbf{Model} & \textbf{Accuracy} & \textbf{F1} & \textbf{Precision} & \textbf{Time} \\
\midrule
DS2OS & RF & 96.26\% & 0.9695 & 0.9830 & 23.48s \\
DS2OS & XGBoost & 96.26\% & 0.9695 & 0.9830 & 3.60s \\
DS2OS & LightGBM & 96.26\% & 0.9695 & 0.9830 & 5.60s \\
DS2OS & Gradient Boosting & 96.23\% & 0.9693 & 0.9829 & 48.36s \\
DS2OS & DT & 96.26\% & 0.9695 & 0.9830 & 0.37s \\
DS2OS & KNN & 99.21\% & 0.9915 & 0.9921 & 1.18s \\
\midrule
NSL-KDD & RF & 77.17\% & 0.7686 & 0.8345 & 9.29s \\
NSL-KDD & XGBoost & 80.47\% & 0.8037 & 0.8511 & 4.01s \\
NSL-KDD & LightGBM & 80.35\% & 0.8023 & 0.8503 & 2.03s \\
NSL-KDD & Gradient Boosting & 78.25\% & 0.7802 & 0.8391 & 47.76s \\
NSL-KDD & DT & 77.66\% & 0.7740 & 0.8354 & 0.98s \\
NSL-KDD & KNN & 76.76\% & 0.7639 & 0.8347 & 0.04s \\
\bottomrule
\end{tabular}
\par\vspace{4pt}
\parbox{\textwidth}{\footnotesize \textit{Note:} DS2OS results use a stratified 70/30 split. NSL-KDD results use the official \mbox{KDDTrain+}/\allowbreak\mbox{KDDTest+} holdout unless otherwise specified. Across 10 seeds the accuracy standard deviation is below 0.2 points on DS2OS and under 1 point on NSL-KDD and the deterministic learners (XGBoost, LightGBM, KNN) show zero seed variance on the fixed NSL-KDD holdout, so the single-split values are stable point estimates.}
\end{table}

On NSL-KDD, ensemble methods such as XGBoost, LightGBM and Gradient Boosting outperformed simpler approaches such as DT and KNN, though all models showed reduced performance when evaluated on novel attack types. The results highlight the substantial accuracy gap between DS2OS (96 to 99 percent) and NSL-KDD (77 to 80 percent).

\subsection{Operational Metrics Beyond Accuracy}
\label{subsec:operational}
The accuracy and weighted F1 in Table~\ref{tab:validation_results} are support-weighted across both classes, so on imbalanced data they are governed by the easy normal class.

To provide a more comprehensive evaluation, Table~\ref{tab:operational} reports attack-class precision, recall, attack F1-score, false positive rate, and Matthews Correlation Coefficient (MCC).

\begin{table}[t]
\caption{Operational attack-class metrics derived from the confusion matrices.}
\label{tab:operational}
\centering
\setlength{\tabcolsep}{5pt}
\begin{tabular}{@{}llccccc@{}}
\toprule
\textbf{Dataset} & \textbf{Model} & \textbf{Prec$_a$} & \textbf{Rec$_a$} & \textbf{F1$_a$} & \textbf{FPR} & \textbf{MCC} \\
\midrule
DS2OS & RF & 0.426 & 0.966 & 0.591 & 0.038 & 0.628 \\
DS2OS & XGBoost & 0.426 & 0.966 & 0.591 & 0.038 & 0.628 \\
DS2OS & LightGBM & 0.426 & 0.966 & 0.591 & 0.038 & 0.628 \\
DS2OS & Gradient Boosting & 0.424 & 0.966 & 0.589 & 0.038 & 0.627 \\
DS2OS & DT & 0.426 & 0.966 & 0.591 & 0.038 & 0.628 \\
DS2OS & KNN & 0.992 & 0.724 & 0.837 & 0.000 & 0.844 \\
\midrule
NSL-KDD & RF & 0.967 & 0.620 & 0.756 & 0.028 & 0.609 \\
NSL-KDD & XGBoost & 0.969 & 0.679 & 0.798 & 0.029 & 0.657 \\
NSL-KDD & LightGBM & 0.968 & 0.677 & 0.797 & 0.029 & 0.655 \\
NSL-KDD & Gradient Boosting & 0.966 & 0.640 & 0.770 & 0.030 & 0.624 \\
NSL-KDD & DT & 0.964 & 0.631 & 0.763 & 0.031 & 0.615 \\
NSL-KDD & KNN & 0.971 & 0.610 & 0.749 & 0.024 & 0.605 \\
\bottomrule
\end{tabular}
\par\vspace{4pt}
\parbox{\textwidth}{\footnotesize Prec$_a$, Rec$_a$ and F1$_a$ are computed with the attack class as positive. The weighted F1 in Table~\ref{tab:validation_results} reconciles exactly with these confusion matrices.}
\end{table}

Three observations follow. First, on DS2OS the tree-based models carry a weighted F1 of about 0.97 yet an attack-class precision of only 0.43, which means roughly 57 percent of the events they flag are false alarms. This is the base-rate fallacy that has long complicated intrusion detection: when attacks are rare, even a low FPR produces alerts that are mostly wrong~\cite{falsePositive}. A model presented as near-perfect raises a false alert on the majority of its detections. Second, KNN inverts the failure rather than resolving it. Its attack precision reaches 0.99, but its recall falls to 0.72, so it misses more than one in four attacks, which is the worse error for a security task where a missed miner runs unchecked. Third, the inflation scales with imbalance: on the severely imbalanced DS2OS the weighted F1 of 0.97 overstates the attack-class F1 of 0.59 by roughly 0.38, while on the near-balanced NSL-KDD the two nearly coincide.

MCC further confirms this observation. No evaluated model exceeds an MCC of 0.844, indicating that overall classifier performance is more modest than suggested by the reported accuracy values.

The apparent gulf between DS2OS at 96 to 99 percent accuracy and NSL-KDD at 77 to 80 percent largely closes once the majority class is removed from the score. With the partial exception of KNN on DS2OS, which reaches 0.844 by trading recall for precision, the configurations describe modest attack-class detectors rather than the near-perfect systems the accuracy numbers imply. This finding compounds the leakage and generalization results. Even after a result is reproduced and even when it is evaluated on familiar threats, the headline F1 can still overstate the only performance that matters operationally.

\subsection{Impact of Feature Leakage}\label{subsec:leak}
\begin{figure}[t]
\centering
\includegraphics[width=0.9\textwidth, height = 5cm]{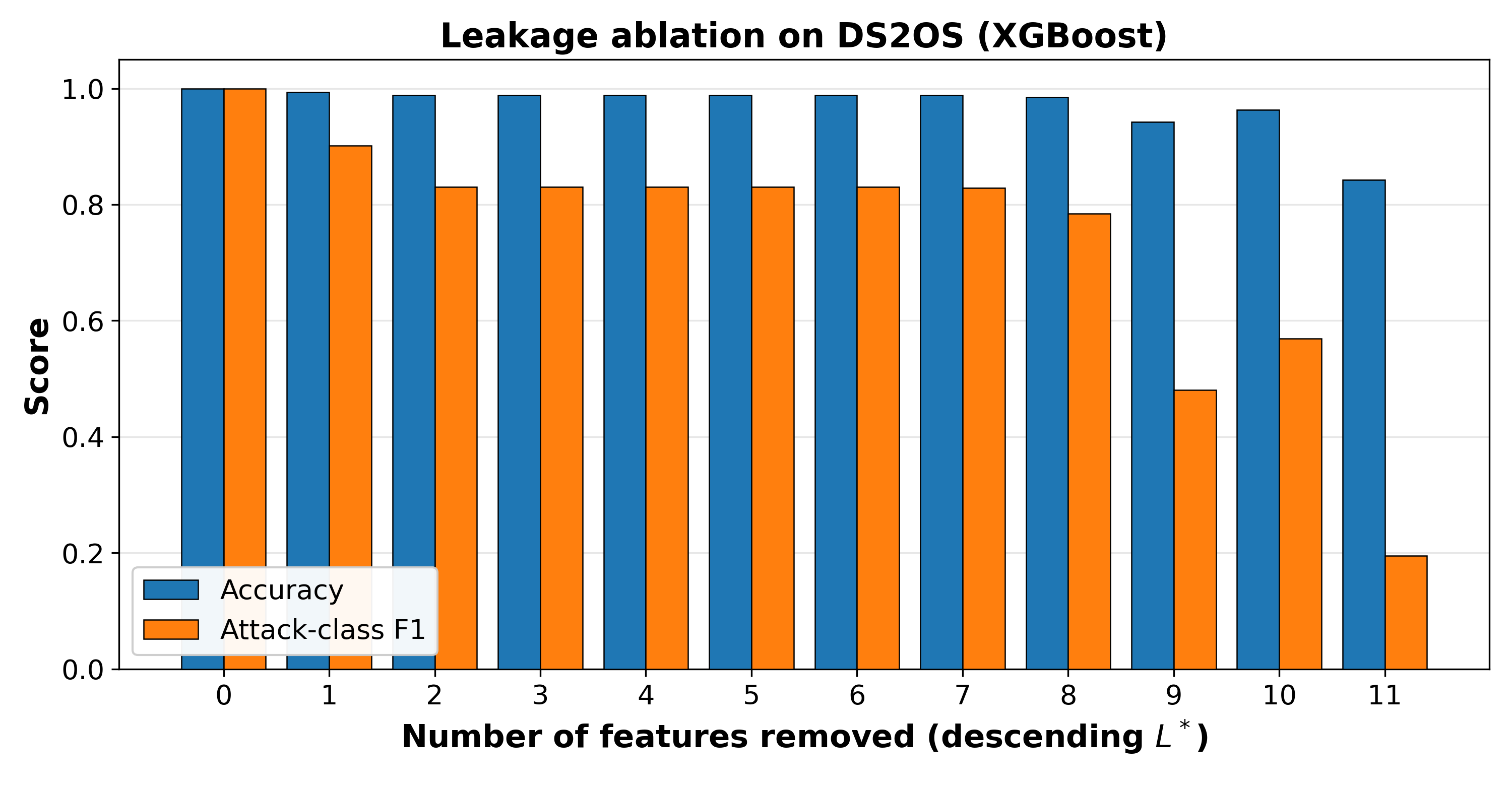}
%\Description{Grouped bar chart of XGBoost accuracy and attack-class F1 on DS2OS, one bar pair for each number of removed features, with features removed one at a time in descending order of adjusted leakage score.}
\caption{Effect of removing DS2OS features in descending order of adjusted leakage score $L^{*}$ (Table~\ref{tab:leakage}) on XGBoost accuracy and attack-class F1. Each bar pair is an independent run of the full pipeline after removing the $k$ highest-ranked features.}
\label{fig:leakage_ablation}
\end{figure}

To evaluate the effect of preprocessing, we compared model performance before and after removing identifier-related features from the DS2OS dataset. With all twelve raw features retained, XGBoost reaches 0.99996 accuracy and 0.999 attack-class F1, which is the near-perfect regime reported in prior DS2OS studies~\cite{ref39,ref21}. The leakage screen explains why. Against a majority base rate of $p_{\mathrm{maj}}=0.972$, the timestamp column scores $L=0.9999$, an adjusted lift $L^{*}$ near 1.0, because attacks were recorded in sequential time blocks that make the timestamp a near-deterministic label proxy. A cluster of address and identifier columns (value, sourceAddress, sourceID, accessedNodeAddress) carries moderate lift, while the location features carry almost none beyond the base rate. Fig.~\ref{fig:leakage_ablation} shows the consequence. Removing the timestamp column alone drops attack-class F1 from 0.999 to 0.90 and removing the remaining identifier and address columns brings it to the low-0.8 range. Among these identifier columns, the largest single drop coincides with timestamp, the highest-lift column, which validates the screen as a ranking and not only as a binary flag. The clean pipeline in Table~\ref{tab:validation_results}, which retains five behavioral features, lands at an attack-class F1 of 0.59. That pipeline is not a step of the ablation, since it keeps three service attributes that rank high on $L^{*}$ and drops accessedNodeType and operation on semantic grounds, as Table~\ref{tab:leakage} explains. The lesson is twofold. The headline accuracy was an artifact of a timestamp leak and on imbalanced data that leak is invisible to a raw single-feature accuracy because every feature clears 0.97 on the base rate alone. A 99 percent result built on a sequential identifier is not a detector. It is a lookup table. The dispositions in Table~\ref{tab:leakage} follow one rule. $L^{*}$ flags near-deterministic leaks such as timestamp, while identifier-typed columns are removed regardless of their measured lift, because a low in-sample lift on a single split is not guaranteed to transfer across collections. Because destinationServiceAddress carries an adjusted lift of 0.333, statistically indistinguishable from the removed accessedNodeAddress at 0.335, we verified its disposition empirically. Removing destinationServiceAddress from the clean pipeline changes the accuracy of the five tree-based models by less than 0.01 points and their attack-class F1 by less than 0.001. KNN is the only model that moves, losing 0.57 accuracy points and 0.046 attack-class F1, so the tree-based conclusions are insensitive to the choice on this boundary case.

\begin{table}[t]
\caption{DS2OS feature leakage, base-rate-adjusted lift ($p_{\mathrm{maj}}=0.972$).}
\label{tab:leakage}
\centering
\setlength{\tabcolsep}{5pt}
\begin{tabular}{@{}lccl@{}}
\toprule
\textbf{Feature} & \textbf{$L(f)$} & \textbf{$L^{*}(f)$} & \textbf{Disposition} \\
\midrule
timestamp & 0.9999 & 0.998 & removed (flagged leak) \\
value & 0.9859 & 0.497 & removed (identifier) \\
sourceAddress & 0.9833 & 0.403 & removed (identifier) \\
sourceID & 0.9827 & 0.383 & removed (identifier) \\
accessedNodeAddress & 0.9814 & 0.335 & removed (identifier) \\
destinationServiceAddress & 0.9813 & 0.333 & retained \\
destinationServiceType & 0.9810 & 0.323 & retained \\
sourceType & 0.9802 & 0.292 & retained \\
accessedNodeType & 0.9727 & 0.025 & removed (identifier) \\
operation & 0.9724 & 0.015 & removed (identifier) \\
sourceLocation & 0.9720 & 0.000 & retained \\
destinationLocation & 0.9720 & 0.000 & retained \\
\bottomrule
\end{tabular}
\par\vspace{4pt}
\parbox{\textwidth}{\footnotesize Raw $L(f)$ is the single-feature lookup accuracy; every feature clears the $0.972$ base rate, so it cannot separate leaks on its own. The adjusted lift $L^{*}(f)$ does: only timestamp meets the $0.95$ bar. Identifier and address fields were removed based on semantic role rather than leakage score alone. We retain service and location attributes because they represent behavioral routing context, whereas source IDs, raw addresses, timestamps and operation identifiers encode record-specific or collection-specific artifacts that are unlikely to transfer across environments. The value column is removed on the same basis because it records the payload of each service call, content specific to individual interactions rather than a behavioral attribute, and its moderate lift indicates it partially memorizes the label through payload regularities.}
\end{table}

\subsection{Evaluation Methodology Divergence}\label{subsec:evaldiv}
To investigate the influence of evaluation methodology, models were evaluated using both a random train-test split and the official NSL-KDD train-test partition. To remove the confound of comparing against externally reported numbers, we ran both protocols through one identical pipeline, with results in Table~\ref{tab:gengap}. A random stratified split, whose test set contains only attack types seen in training, produces substantially higher accuracy because all attack types are represented during training. The official \mbox{KDDTest+} holdout, which adds novel attack types, drops the same models on the same pipeline to 0.77 to 0.80. The generalization gap $G$, defined as the difference between seen-attack and novel-attack accuracy, is 0.19 to 0.23. Because both numbers come from one pipeline, the gap is attributable to the seen-versus-novel protocol alone, not to differences in preprocessing or features. This is the central operational risk for cryptojacking detection, where attackers continuously introduce evasion techniques that a deployed model must handle without retraining. The same bias appears in malware classification, where evaluation that ignores temporal and family novelty has been shown to overstate effectiveness~\cite{pendlebury2019tesseract}.

\begin{table}[t]
\caption{Performance under different evaluation protocols, identical pipeline.}
\label{tab:gengap}
\centering
\setlength{\tabcolsep}{5pt}
\begin{tabular}{@{}lccc@{}}
\toprule
\textbf{Model} & \textbf{Acc$_{\text{seen}}$ (random)} & \textbf{Acc$_{\text{novel}}$ (holdout)} & \textbf{$G$} \\
\midrule
KNN & 0.996 & 0.768 & 0.228 \\
RF & 0.999 & 0.772 & 0.227 \\
DT & 0.998 & 0.777 & 0.221 \\
Gradient Boosting & 0.999 & 0.782 & 0.216 \\
LightGBM & 0.999 & 0.803 & 0.196 \\
XGBoost & 0.999 & 0.805 & 0.194 \\
\bottomrule
\end{tabular}
\par\vspace{4pt}
\parbox{\textwidth}{\footnotesize Acc$_{\text{seen}}$ is a stratified 70/30 random split of \mbox{KDDTrain+} that exposes all attack types in training. Acc$_{\text{novel}}$ is the official \mbox{KDDTest+} holdout and matches Table~\ref{tab:validation_results}. Both use the same pipeline and seed, so $G$ isolates the seen-versus-novel effect. $G$ is computed from unrounded accuracies.}
\end{table}

\begin{figure}[t]
\centering
\includegraphics[width=\textwidth,height=5.25cm ]{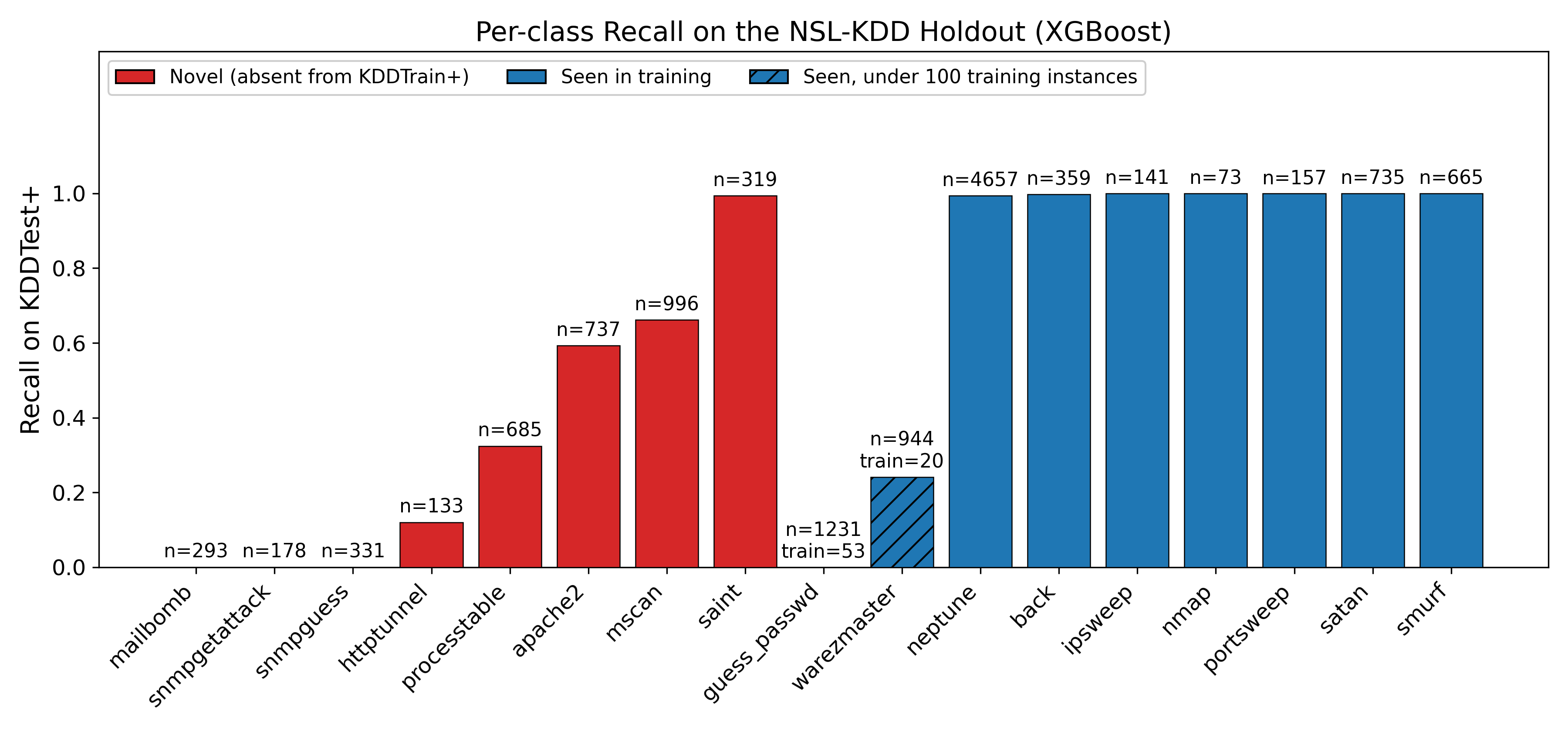}
%\Description{Bar chart of XGBoost recall on each NSL-KDD test attack family with at least 50 samples. Novel families absent from training appear in red: mailbomb, snmpgetattack and snmpguess at zero recall, httptunnel at 0.12, processtable at 0.32, apache2 at 0.59, mscan at 0.66 and saint at 0.99. Two rarely seen families appear hatched: guess\_passwd at zero and warezmaster at 0.24. Well-represented seen families in blue, such as neptune, back, ipsweep, nmap, portsweep, satan and smurf, reach recall of 0.99 to 1.0.}
\caption{Per-class recall of XGBoost on the official NSL-KDD holdout for attack families with at least 50 test samples. Blue bars represent attack families observed during training, red bars represent previously unseen attack families, and hatched bars indicate attack families with fewer than 100 training instances. Sample counts are shown above each bar.}
\label{fig:per_class_recall}
\end{figure}

To show where the gap originates, Fig.~\ref{fig:per_class_recall} reports recall on each \mbox{KDDTest+} attack type. The failures concentrate in families the model has little or no representation for. Several substantial novel attacks are missed outright: mailbomb at 0.00 recall over 293 samples, snmpguess at 0.00 over 331 and snmpgetattack at 0.00 over 178. Others are caught only partially, such as processtable at 0.32, apache2 at 0.59 and mscan at 0.66, while saint is the exception among novel families at 0.99, presumably because its probing behavior resembles the heavily represented satan. Two seen families sharpen rather than contradict this picture. The guess\_passwd family appears in \mbox{KDDTrain+} with only 53 instances yet contributes 1,231 holdout samples at zero recall. The warezmaster family, with 20 training instances, is recovered at 0.24. Well-represented seen families such as neptune at 0.99, satan at 1.00 and smurf at 1.00 are detected almost perfectly. The aggregate gap is therefore driven by training representation: families that are novel or vanishingly rare in training fail, which is exactly the failure a deployed detector must avoid and the reason detection models age and require drift-aware monitoring once deployed~\cite{jordaney2017transcend}.

\subsection{Computational Cost}
The training times in Table~\ref{tab:validation_results} show the trade-off between cost and accuracy. On DS2OS, DT trained in 0.37 seconds while matching the accuracy of heavier ensembles. XGBoost and LightGBM trained in 3.60 and 5.60 seconds with the same accuracy that RF reached in 23.48 seconds. Gradient Boosting needed 48.36 seconds for marginally lower accuracy. On NSL-KDD, XGBoost delivered the best accuracy of 80.47 percent in 4.01 seconds, while Gradient Boosting needed 47.76 seconds for a lower 78.25 percent. The implication is consistent across datasets. XGBoost and LightGBM offer the best accuracy-to-cost ratio and suit online learning and continuous retraining. Heavier sequential boosting offers marginal gains at large cost. KNN is excluded from this comparison because it defers computation to inference. Its near-zero training time reflects instance storage rather than model fitting and its per-query cost grows with the size of the training set, which makes it the least suitable candidate for online retraining despite its apparent speed in Table~\ref{tab:validation_results}.

\subsection{Class Imbalance Behavior}
Without SMOTE, preliminary DS2OS models reached 97 percent accuracy by predicting the majority class for every sample, with zero attack recall. SMOTE generated 236,542 synthetic attack samples for training only. On the untouched test set of 104,381 normal and 3,005 attack samples, the tree-based models missed only 101 attacks but raised 3,915 to 3,945 false alarms. KNN raised only 17 false alarms yet missed 828 attacks. On NSL-KDD, RF missed 4,875 attacks, predominantly from the novel families absent from training, while XGBoost missed the fewest at 4,119.

%%%%%%%%%%%%%%%%%%%%%%%%%%%%%%%%%%%%%%%%%%%%%%%%%%%%%%%%%%%%%%%%%%%%%%%%%%%%%%%
\section{Discussion}
\label{sec:discussion}
The empirical evaluation presented in this study demonstrates that the performance of classical machine learning models for cloud cryptojacking detection is influenced not only by the learning algorithm but also by preprocessing choices, dataset characteristics, and evaluation protocols. Across both datasets, the evaluated models exhibited different trade-offs between overall accuracy, attack detection capability, and computational efficiency.

Because no publicly available cloud-native cryptojacking dataset currently exists, this study relied on two publicly available proxy datasets. Although DS2OS and NSL-KDD do not capture cloud-native telemetry directly, they provide complementary evaluation scenarios for comparing machine learning models under consistent experimental conditions. Consequently, the reported results should be interpreted as empirical observations on these datasets rather than direct indicators of operational cloud performance.

Removing identifier-related features from DS2OS reduced the performance of most evaluated models, demonstrating that preprocessing choices can substantially influence reported classifier performance. These findings highlight the importance of documenting preprocessing procedures when comparing machine learning results across studies.

The NSL-KDD experiments showed that evaluation protocol has a substantial impact on observed model performance. Accuracy was considerably higher under a random train-test split than under the official holdout containing previously unseen attack types, indicating that different evaluation protocols can lead to substantially different performance estimates.

Cross-environment transfer fails at the feature level. A direct attempt to reuse a model across the two datasets is impossible by construction, since DS2OS retains five behavioral features while NSL-KDD has 41, so portability is a fundamental technical barrier rather than a convenience.
Future work may investigate whether domain adaptation or feature alignment techniques can improve cross-dataset transferability.

The results also demonstrate the importance of reporting complementary evaluation metrics. While overall accuracy remained high for several models, attack-class precision, recall, and Matthews Correlation Coefficient (MCC) provided additional insight into classifier behavior, particularly for the imbalanced DS2OS dataset.

Overall, this study provides a consistent empirical comparison of six classical machine learning models under common preprocessing and evaluation settings. The results demonstrate that dataset characteristics, preprocessing strategies, and evaluation protocols all influence observed model performance. Future work should extend this empirical evaluation to publicly available cloud-native cryptojacking datasets and include additional machine learning approaches as such datasets become available.

%%%%%%%%%%%%%%%%%%%%%%%%%%%%%%%%%%%%%%%%%%%%%%%%%%%%%%%%%%%%%%%%%%%%%%%%%%%%%%%
\section{Threats to Validity}
\label{sec:threats}

This study has several limitations. First, DS2OS and NSL-KDD are proxy datasets rather than cloud-native cryptojacking datasets, so the reported results may not directly reflect performance in operational cloud environments. Second, only classical machine learning models were evaluated; deep learning and hybrid approaches were outside the scope of this work. Third, hyperparameters were selected from commonly reported configurations rather than being individually optimized for each model. Fourth, although repeated experiments demonstrated stable results across random seeds, additional validation using other public datasets would further strengthen the generality of the findings.

Finally, label encoding introduces an arbitrary ordering of categorical features, which may particularly influence the distance calculations used by K-Nearest Neighbors (KNN). However, the SMOTE-N sensitivity analysis presented in Section~\ref{subsec:preprocessing} showed only negligible differences in performance, suggesting that this effect is limited.

The use of two proxy datasets and six classical machine learning models also limits the generalizability of the reported performance. Future studies should extend this empirical evaluation to additional public datasets and a broader range of machine learning approaches, including deep learning models.

%%%%%%%%%%%%%%%%%%%%%%%%%%%%%%%%%%%%%%%%%%%%%%%%%%%%%%%%%%%%%%%%%%%%%%%%%%%%%%%
\section{Conclusion}
\label{sec:conclusion}
This paper presented an empirical evaluation of six widely used classical machine learning models for AI-based cloud cryptojacking detection using two publicly available proxy datasets, DS2OS and NSL-KDD. The evaluation examined the influence of preprocessing choices, dataset characteristics, and evaluation protocols on model performance under consistent experimental settings.

The experimental results showed that model performance varies considerably depending on the evaluation scenario. On DS2OS, removing identifier-related features reduced the performance of most evaluated models, demonstrating the impact of preprocessing on reported results. On NSL-KDD, the official train-test partition produced substantially lower accuracy than a random train-test split, highlighting the influence of evaluation protocol on observed classifier performance. In addition, attack-class precision, recall, and Matthews Correlation Coefficient (MCC) complemented overall accuracy by providing a more comprehensive assessment of model behavior, particularly for imbalanced datasets.

Overall, this study provides a consistent empirical comparison of representative classical machine learning models for cloud cryptojacking detection and offers practical insights into how preprocessing strategies and evaluation methodology affect reported performance. Future work will extend this evaluation to publicly available cloud-native cryptojacking datasets, additional machine learning models, and more realistic cloud computing environments as such datasets become available.

\begin{credits}
\subsubsection{\discintname}
The authors have no competing interests to declare that are relevant to the content of this article.
\end{credits}
\bibliographystyle{unsrt}
\bibliography{references}

\begin{thebibliography}{10}
\providecommand{\url}[1]{\texttt{#1}}
\providecommand{\urlprefix}{URL }
\providecommand{\doi}[1]{https://doi.org/#1}

\bibitem{ref27}
Ahsan, M.A., Arshad, A., Mian, A.N.: Leveraging tabular {GANs} for malicious
  address classification in {Ethereum} network. Computer Networks
  \textbf{254},  110813 (2024). \doi{10.1016/j.comnet.2024.110813}

\bibitem{arp2022dos}
Arp, D., Quiring, E., Pendlebury, F., Warnecke, A., Pierazzi, F., Wressnegger,
  C., Cavallaro, L., Rieck, K.: Dos and don'ts of machine learning in computer
  security. In: 31st USENIX Security Symposium (USENIX Security 22). pp.
  3971--3988 (2022),
  \url{https://www.usenix.org/conference/usenixsecurity22/presentation/arp}

\bibitem{ds2os_dataset}
Aubet, F.X., Pahl, M.O.: {DS2OS} traffic traces. Kaggle (2018),
  \url{https://www.kaggle.com/datasets/francoisxa/ds2ostraffictraces}, accessed
  in 2026; alternative Kaggle mirror:
  https://www.kaggle.com/datasets/libamariyam/ds2os-dataset

\bibitem{falsePositive}
Axelsson, S.: The base-rate fallacy and the difficulty of intrusion detection.
  ACM Trans. Inf. Syst. Secur.  \textbf{3}(3),  186--205 (2000).
  \doi{10.1145/357830.357849}

\bibitem{ref25}
Bhardwaj, S., Li, A.S., Dave, M., Bertino, E.: Overcoming the lack of labeled
  data: Training malware detection models using adversarial domain adaptation.
  Computers \& Security  \textbf{140},  103769 (2024).
  \doi{10.1016/j.cose.2024.103769}

\bibitem{nslkdd_dataset}
{Canadian Institute for Cybersecurity}: {NSL-KDD} dataset. University of New
  Brunswick (2009), \url{https://www.unb.ca/cic/datasets/nsl.html}, accessed in
  2026; Kaggle mirror: https://www.kaggle.com/datasets/hassan06/nslkdd

\bibitem{chakravorty2026code}
Chakravorty, A., Elsayed, N.: Code for ``{Empirical Evaluation of Machine
  Learning for Cloud Cryptojacking Detection}''. GitHub repository (2026),
  \url{https://github.com/AmitabhCh822/cloud-cryptojacking-ml-evaluation}

\bibitem{chakravorty2026detecting}
Chakravorty, A., Elsayed, N.: Detecting cryptojacking in cloud environments: A
  systematic review of ai-based defenses, deployment challenges, and research
  gaps (2026), manuscript submitted for publication

\bibitem{chawla2002smote}
Chawla, N.V., Bowyer, K.W., Hall, L.O., Kegelmeyer, W.P.: {SMOTE}: Synthetic
  minority over-sampling technique. Journal of Artificial Intelligence Research
   \textbf{16},  321--357 (2002). \doi{10.1613/jair.953}

\bibitem{chicco2020mcc}
Chicco, D., Jurman, G.: The advantages of the {Matthews} correlation
  coefficient ({MCC}) over {F1} score and accuracy in binary classification
  evaluation. BMC Genomics  \textbf{21}(1), ~6 (2020).
  \doi{10.1186/s12864-019-6413-7}

\bibitem{jordaney2017transcend}
Jordaney, R., Sharad, K., Dash, S.K., Wang, Z., Papini, D., Nouretdinov, I.,
  Cavallaro, L.: Transcend: Detecting concept drift in malware classification
  models. In: 26th USENIX Security Symposium (USENIX Security 17). pp. 625--642
  (2017),
  \url{https://www.usenix.org/conference/usenixsecurity17/technical-sessions/presentation/jordaney}

\bibitem{kapoor2023leakage}
Kapoor, S., Narayanan, A.: {Leakage and the reproducibility crisis in
  machine-learning-based science}. Patterns  \textbf{4}(9),  100804 (2023).
  \doi{10.1016/j.patter.2023.100804}

\bibitem{kaufman2012leakage}
Kaufman, S., Rosset, S., Perlich, C., Stitelman, O.: Leakage in data mining:
  Formulation, detection, and avoidance. ACM Trans. Knowl. Discov. Data
  \textbf{6}(4),  1--21 (2012). \doi{10.1145/2382577.2382579}

\bibitem{ref10}
Kim, R., Ryu, J., Kim, S., Lee, S., Kim, S.: Detecting cryptojacking containers
  using {eBPF}-based security runtime and machine learning. Electronics
  \textbf{14}(6), ~1208 (2025). \doi{10.3390/electronics14061208}

\bibitem{ref7}
Kwedza, P., Chindipha, S.D.: Cryptojacking detection in cloud infrastructure
  using network traffic. In: 2023 International Conference on Electrical,
  Computer and Energy Technologies (ICECET). pp.~1--6 (2023).
  \doi{10.1109/ICECET58911.2023.10389593}

\bibitem{ref20}
Lin, Y.D., Yang, S.Y., Sudyana, D., Yudha, F., Lai, Y.C., Hwang, R.H.:
  Two-stage multi-datasource machine learning for attack technique and
  lifecycle detection. Computers \& Security  \textbf{142},  103859 (2024).
  \doi{10.1016/j.cose.2024.103859}

\bibitem{microsoft2023cryptojacking}
{Microsoft Security Blog}: Cryptojacking: Understanding and defending against
  cloud compute resource abuse (2023),
  \url{https://www.microsoft.com/en-us/security/blog/2023/07/25/cryptojacking-understanding-and-defending-against-cloud-compute-resource-abuse/},
  accessed: 2025

\bibitem{naseem2021minos}
Naseem, F., Aris, A., Babun, L., Tekiner, E., Uluagac, A.S.: {MINOS}: A
  lightweight real-time cryptojacking detection system. In: Network and
  Distributed System Security Symposium (NDSS) (2021).
  \doi{10.14722/ndss.2021.24444}

\bibitem{pendlebury2019tesseract}
Pendlebury, F., Pierazzi, F., Jordaney, R., Kinder, J., Cavallaro, L.:
  {TESSERACT}: Eliminating experimental bias in malware classification across
  space and time. In: 28th USENIX Security Symposium (USENIX Security 19). pp.
  729--746 (2019),
  \url{https://www.usenix.org/conference/usenixsecurity19/presentation/pendlebury}

\bibitem{sanda2023deep}
Sanda, O., Pavlidis, M., Polatidis, N.: A deep learning approach for host-based
  cryptojacking malware detection. Evolving Systems  \textbf{15}(1),  41--56
  (2024). \doi{10.1007/s12530-023-09534-9}

\bibitem{ref32}
Sethuraman, S.C., Jadapalli, T.G., Sudhakaran, D.P.V., Mohanty, S.P.: Flow
  based containerized honeypot approach for network traffic analysis: An
  empirical study. Computer Science Review  \textbf{50},  100600 (2023).
  \doi{10.1016/j.cosrev.2023.100600}

\bibitem{ref29}
Sun, P., Lyu, M., Li, H., Yang, B., Peng, L.: An early stage convolutional
  feature extracting method using for mining traffic detection. Computer
  Communications  \textbf{193},  346--354 (2022).
  \doi{10.1016/j.comcom.2022.06.044}

\bibitem{tavallaee2009nslkdd}
Tavallaee, M., Bagheri, E., Lu, W., Ghorbani, A.A.: A detailed analysis of the
  {KDD CUP} 99 data set. In: 2009 IEEE Symposium on Computational Intelligence
  for Security and Defense Applications. pp.~1--6 (2009).
  \doi{10.1109/CISDA.2009.5356528}

\bibitem{ref39}
Tekin, N., Acar, A., Aris, A., Uluagac, A.S., Gungor, V.C.: Energy consumption
  of on-device machine learning models for {IoT} intrusion detection. Internet
  of Things  \textbf{21},  100670 (2023). \doi{10.1016/j.iot.2022.100670}

\bibitem{tekiner2021sok}
Tekiner, E., Acar, A., Uluagac, A.S., Kirda, E., Selcuk, A.A.: Sok:
  cryptojacking malware. In: 2021 IEEE European Symposium on Security and
  Privacy (EuroS\&P). pp. 120--139. IEEE (2021)

\bibitem{ref21}
Tiwari, D., Bhati, B.S., Nagpal, B., Sankhwar, S., Al-Turjman, F.: An enhanced
  intelligent model: To protect marine {IoT} sensor environment using ensemble
  machine learning approach. Ocean Engineering  \textbf{242},  110180 (2021).
  \doi{10.1016/j.oceaneng.2021.110180}

\bibitem{varlioglu2022dangerous}
Varlioglu, S., Elsayed, N., ElSayed, Z., Ozer, M.: The dangerous combo:
  Fileless malware and cryptojacking. In: SoutheastCon 2022. pp. 125--132. IEEE
  (2022)

\bibitem{varlioglu2026detecting}
Varlioglu, S., Elsayed, N., Ozer, M., ElSayed, Z., Emmert, J.M.: Detecting
  fileless cryptojacking in {PowerShell} using {AST}-enhanced {CodeBERT}
  models. arXiv preprint, arXiv:2602.18285 (2026),
  \url{https://arxiv.org/abs/2602.18285}

\bibitem{varlioglu2024pulse}
Varlioglu, S., Elsayed, N., Varlioglu, E.R., Ozer, M., ElSayed, Z.: The pulse
  of fileless cryptojacking attacks: malicious {PowerShell} scripts. In:
  SoutheastCon 2024. pp. 571--580. IEEE (2024)

\end{thebibliography}
\end{document}